\documentclass[11pt]{article}
\PassOptionsToPackage{authoryear}{natbib}
\usepackage[preprint,nonatbib]{neurips_2024}
\usepackage{hyperref} 
\usepackage{authblk} 
\usepackage{booktabs}
\usepackage{graphicx}
\usepackage{amsfonts}
\usepackage{multicol}
\usepackage{tabularx}
\usepackage{enumitem,amsmath,bm,amssymb}
\usepackage{pifont,makecell}
\usepackage{tikz}
\usepackage{CJKutf8}

\usepackage{multirow}
\usepackage{cite}
\usepackage{listings}
\usepackage{caption} 
\title{Muyan-TTS: A Trainable Text-to-Speech Model Optimized for Podcast Scenarios with a \$50K Budget}

\author{
  \bf Xin Li, Kaikai Jia, Hao Sun, Jun Dai, Ziyang Jiang
}

\affil{MYZY AI}
\affil{\texttt{kearney@myzy.ai}}
\date{}
\begin{document}

\maketitle

\begin{abstract}
Recent advancements in text-to-speech (TTS) models have been driven by the integration of large language models (LLMs), enhancing semantic comprehension and improving speech naturalness. However, existing LLM-based TTS models often lack open-source training code and efficient inference acceleration frameworks, limiting their accessibility and adaptability. Additionally, there is no publicly available TTS model specifically optimized for podcast scenarios, which are in high demand for voice interaction applications. To address these limitations, we introduce \textbf{Muyan-TTS}, an open-source trainable TTS model designed for podcast applications within a \$50,000 budget. Our model is pre-trained on over 100,000 hours of podcast audio data, enabling zero-shot TTS synthesis with high-quality voice generation. Furthermore, Muyan-TTS supports speaker adaptation with dozens of minutes of target speech, making it highly customizable for individual voices. In addition to open-sourcing the model, we provide a comprehensive data collection and processing pipeline, a full training procedure, and an optimized inference framework that accelerates LLM-based TTS synthesis. Our code and models are available at \url{https://github.com/MYZY-AI/Muyan-TTS}.
\end{abstract}

\section{Introduction}

In recent years, text-to-speech (TTS) technology has undergone significant advancements, evolving from traditional end-to-end models~\cite{kim2021conditional,kong2023vits2,weiss2021wave,donahue2020end,ping2018clarinet} to more sophisticated cascade models~\cite{gptsovits,chattts,li2019neural,shen2018natural,valle2020flowtron}, where an auto-regressive (AR) model and a decoder can be trained separately. A particularly notable development in this transition is the integration of large language models (LLMs) into the TTS pipeline~\cite{wang2023neural,chen2024vall,du2023lauragpt,liao2024fish,anastassiou2024seed,du2024cosyvoice1,neekhara2024improving,xue2024improving}. Recent works such as CosyVoice 2~\cite{du2024cosyvoice} and Step-Audio~\cite{huang2025step} leverage pre-trained LLMs to enhance the semantic comprehension ability of TTS models, resulting in more natural and contextually coherent speech synthesis. This paradigm shift has enabled TTS models to capture linguistic nuances more effectively, bridging the gap between text processing and speech generation.

Despite these advancements, the accessibility and openness of LLM-based TTS models remain limited. Over the past year, the AI community has seen a surge in open-source contributions to TTS models. However, many LLM-based TTS models do not provide training code or an efficient inference acceleration framework in their official repositories. On the other hand, while GPT-SoVITS~\cite{gptsovits} offers training code for both the AR model and the decoder, the lack of a pre-trained LLM significantly affects the performance of the synthesized speech. Furthermore, while TTS models have demonstrated remarkable progress in various applications, there is still no widely available open-source TTS model specifically designed for podcast scenarios. Given the growing demand for high-quality, natural-sounding voice synthesis in interactive voice applications, the absence of a specialized model optimized for this use case presents a critical gap in the field.

To address these limitations, we introduce Muyan-TTS, a trainable TTS model optimized for podcast scenarios while operating within a \$50,000 budget. Muyan-TTS is built upon the Llama-3.2-3B~\cite{grattafiori2024llama} model, which we continue to pre-train on over 100,000 hours of podcast audio data. The resulting pre-trained model is capable of zero-shot TTS synthesis, producing high-quality speech without requiring extensive fine-tuning. Additionally, Muyan-TTS can be further adapted using dozens of minutes of a target speaker’s voice to enhance its speech synthesis capabilities, making it a highly versatile and customizable TTS solution. Importantly, we present the complete training procedure and open-source the code, ensuring transparency and accessibility to the broader research and developer community. On the decoder side, while cascade models are widely used for TTS, end-to-end models such as VITS~\cite{kim2021conditional,kong2023vits2} have demonstrated superior robustness, particularly in reducing hallucinations due to their grapheme-to-phoneme (G2P) characteristics. To balance efficiency and quality, we integrate an LLM with a VITS-based model, mitigating hallucinations while preserving the advantages of LLM-based TTS synthesis.

This paper makes the following key contributions:
\begin{itemize}[leftmargin=*]
\item Open-sourcing two TTS models: (i) a base model pre-trained on diverse podcast datasets, enabling zero-shot TTS synthesis, and (ii) a supervised fine-tuned (SFT) model trained on an individual speaker to enhance TTS performance.
\item Providing a detailed training methodology: We outline the end-to-end training procedure, from the base model to speaker-specific adaptation, and release the full training code for public use.
\item Introducing a data processing pipeline: We propose a comprehensive workflow for data collection, preprocessing, and formatting tailored to TTS model training, improving efficiency and reproducibility.
\item Optimizing inference efficiency: We develop an accelerated TTS inference framework, particularly optimizing the LLM component for faster and more efficient speech generation.
\end{itemize}

By addressing the current limitations in LLM-based TTS models and introducing an open-source solution tailored for podcast scenarios, Muyan-TTS contributes to the ongoing advancement of speech synthesis technology, making high-quality TTS more accessible and adaptable for real-world applications.

\section{Related Work}

Parallel end-to-end TTS models have demonstrated their ability to synthesize natural-sounding speech waveforms directly from text, bypassing the need for separate linguistic and acoustic modeling stages. A notable example is Variational Inference Text-to-Speech (VITS)~\cite{kim2021conditional}, which leverages a variational autoencoder (VAE) to maximize the variational lower bound and employs adversarial training to enhance synthesis quality. VITS integrates a normalizing flow-based decoder for waveform generation and learns latent representations that improve the expressiveness of generated speech.
An extension of VITS, VITS2~\cite{kong2023vits2}, further enhances synthesis quality by introducing adversarial learning to the duration predictor and employing Monotonic Alignment Search (MAS) for improved naturalness and efficiency. The MAS method allows for better alignment learning, reducing artifacts, and improving prosody modeling. Despite the growing prominence of LLM-based TTS models, VITS remains relevant due to its robustness and reduced hallucination, a known challenge in LLM-based systems. Given this advantage, our work incorporates a VITS-based model to ensure stability in speech synthesis.

Cascade, or two-stage, TTS models have gained traction due to their flexibility in handling different components separately. These models typically decompose the TTS pipeline into two stages: (1) an autoregressive (AR) model that converts text into intermediate representations (e.g., semantic or linguistic tokens), and (2) a decoder model that transforms these representations into speech waveforms. 
A representative cascade model is GPT-SoVITS~\cite{gptsovits}, which first trains an autoregressive model to generate audio semantic tokens from text, followed by a SoVITS model that synthesizes waveforms from these tokens. A major advantage of cascade models is that their components can be trained independently with different datasets and settings, enabling optimization for different aspects of the TTS pipeline. Inspired by this approach, we leverage semantic token extraction from the quantizer in the SoVITS model pre-trained in GPT-SoVITS and subsequently train an LLM for text-to-semantic token conversion.

The integration of pre-trained LLMs into TTS has emerged as a promising approach, leveraging the semantic comprehension capabilities of LLMs to improve speech synthesis. Unlike conventional models, LLM-based TTS architectures benefit from rich contextual understanding, allowing for better handling of prosody, emotion, and speaker characteristics.
One such model, CosyVoice 2~\cite{du2024cosyvoice}, adopts Qwen2.5-0.5B~\cite{yang2024qwen2} as the foundation model, utilizing flow matching and a vocoder for high-quality speech generation. CosyVoice 2 introduces finite-scalar quantization to enhance codebook utilization and implements streaming inference for efficiency. Similarly, Step-Audio~\cite{huang2025step}, built upon Step-1, employs two distinct tokenizers—linguistic and semantic—to encode both semantic and coarse-grained acoustic information, improving overall speech synthesis quality.
Building on these advancements, our work incorporates Llama-3.2-3B as the backbone model for TTS. We detail the training process for both our base TTS model and supervised fine-tuning (SFT) model, along with an efficient inference framework designed to balance quality and computational cost.

By integrating elements from end-to-end, cascade, and LLM-based approaches, our research aims to bridge the gap between robustness and expressiveness in TTS, improving both synthesis quality and inference efficiency.

\section{Methods}

\subsection{Framework}

The architecture of Muyan-TTS builds upon GPT-SoVITS, incorporating a pre-trained large language model (Llama-3.2-3B) as a replacement for the AR model while independently modeling both the LLM and SoVITS components. This approach leverages the strengths of LLMs in text representation while maintaining a robust audio synthesis pipeline.

\begin{figure*}[thb]
	\centering
	\includegraphics[width=\linewidth]{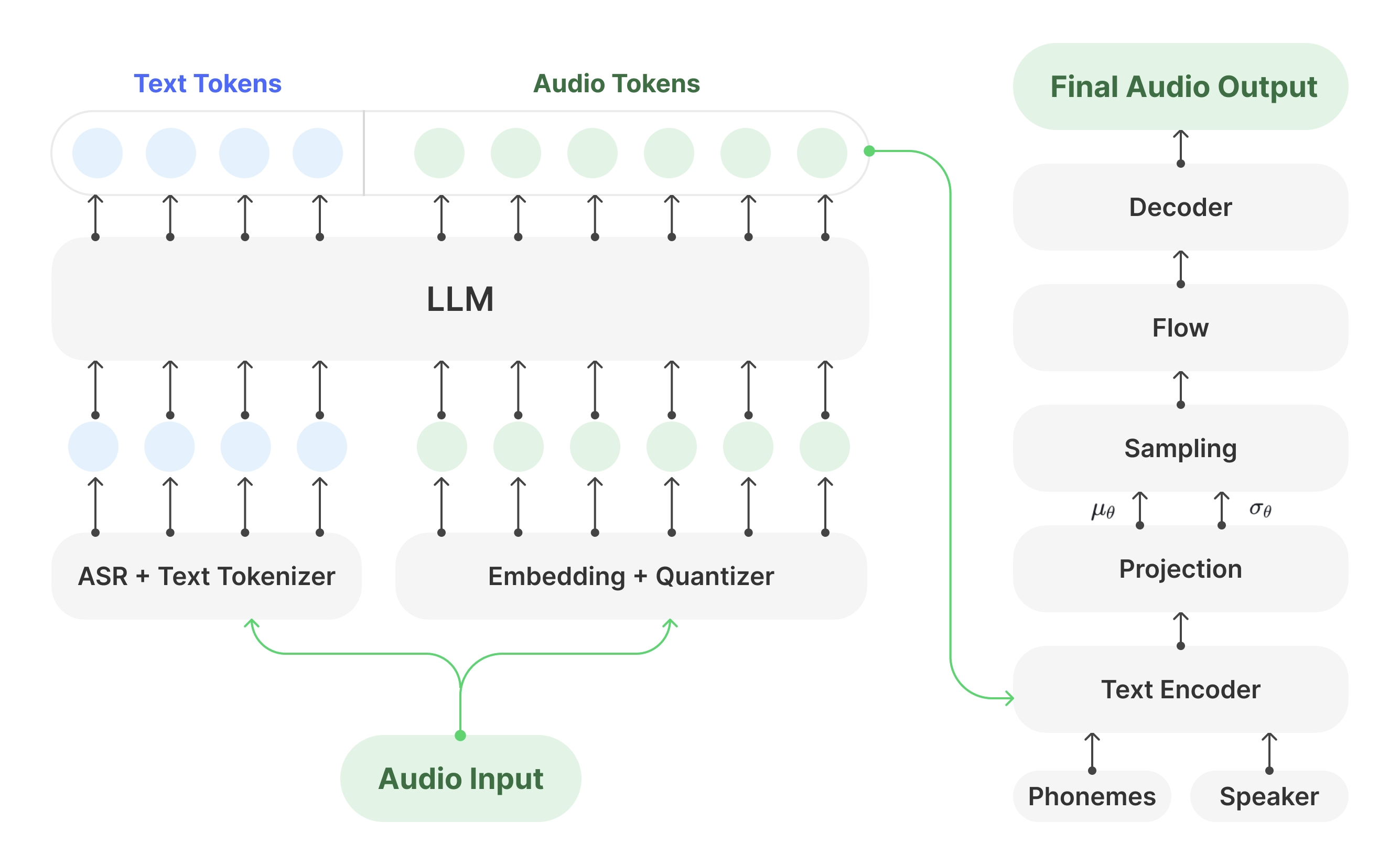}
	\vspace{-0.6cm}
	\caption{Framework of Muyan-TTS. Left is an LLM that models the parallel corpus of text (in blue) and audio (in green) tokens. Right is a SoVITS model that decodes the generated audio tokens, as well as phonemes and speaker embeddings, into the audio waveform.}
	\label{fig:framework}
\end{figure*}

To establish a parallel corpus of text and audio tokens, Muyan-TTS employs an automatic speech recognition (ASR) model, such as Whisper~\cite{radford2023robust}, to transcribe the given audio input, which is then tokenized using the tokenizer of LLM. In parallel, the system extracts Hubert~\cite{hsu2021hubert} embeddings from the audio input with a token rate of 25Hz, which are subsequently quantized using the pre-trained quantizer from GPT-SoVITS, producing discrete audio tokens. By injecting this finite set of audio tokens into the LLM’s vocabulary, Muyan-TTS effectively aligns textual and speech representations within a unified modeling space.

Rather than employing a flow matching model as the decoder, Muyan-TTS adopts a VITS-based model to mitigate hallucinations introduced by the LLM. The VITS model is advantageous due to its G2P nature, which provides a more structured approach to phonetic modeling, thereby enhancing pronunciation accuracy. To further refine the quality of generated speech, we continue training the SoVITS model of GPT-SoVITS on a large-scale high-quality speech dataset. This improves the fidelity and expressiveness of the synthesized audio while maintaining a stable and coherent speech synthesis pipeline.

A high-level overview of the Muyan-TTS framework is illustrated in Figure~\ref{fig:framework}.

\subsection{Datasets}

\begin{figure*}[thb]
	\centering
	\includegraphics[width=\linewidth]{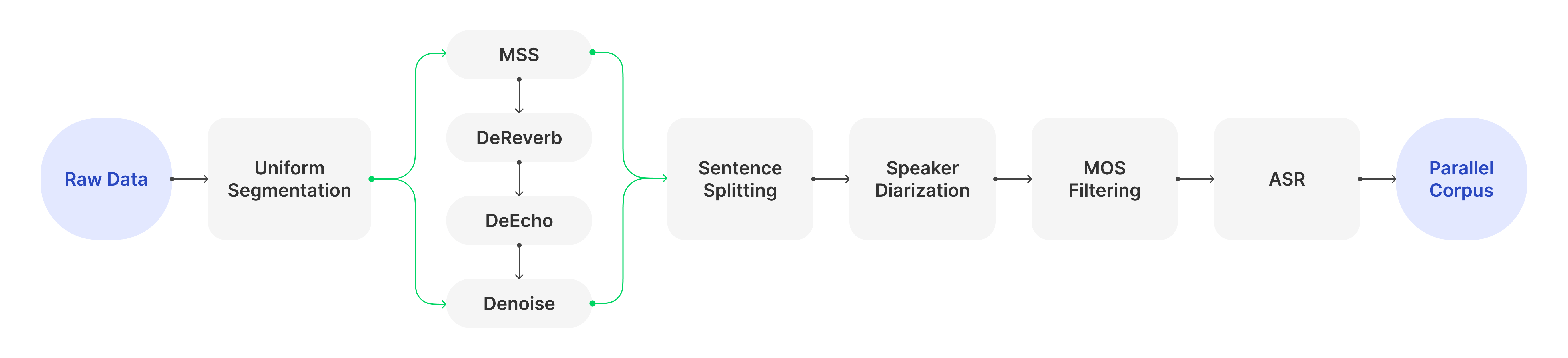}
	\caption{Data processing pipeline.}
	\label{fig:pipeline}
\end{figure*}

To train a high-quality TTS model tailored for podcast scenarios, we construct a large-scale and clean dataset through a carefully designed multi-stage data processing pipeline. An overview of the pipeline is shown in Figure~\ref{fig:pipeline}. The overall process consists of three main stages: data collection, data cleaning, and data formatting.

\textbf{Data Collection.}
To ensure speaker diversity and high-quality speech samples, we collect audio data from both open-source datasets and a proprietary podcast collection. We aim to maximize speaker coverage and linguistic variability, with a focus on English and partial inclusion of Chinese content for increased robustness. Each audio source is uniformly sampled and evaluated using NISQA~\cite{mittag2021nisqa}, a neural network-based speech quality metric that simultaneously estimates intelligibility, naturalness, and overall quality. Based on these scores, we determine the proportion of audio retained from each source.
To facilitate efficient downstream processing, all audio is uniformly segmented into 1-minute chunks, without altering the original format or sample rate. In total, this process yields over 150,000 hours of raw speech data, primarily from English-language podcasts, with a minor share of Chinese-language content.

\textbf{Data Cleaning.}
We apply a comprehensive data cleaning pipeline to enhance the quality of the collected audio. The steps include:
\begin{itemize}[leftmargin=*]
\item MSS (Music Source Separation)~\cite{lu2024music}: separates audio into distinct components such as vocals, bass, and drums. We retain only the vocal tracks and store them in MP3 format.
\item DeReverb~\cite{liu2021voicefixer}: removes reverberations and time delays to reduce spatial artifacts and improve vocal clarity.
\item DeEcho~\cite{yang2023low}: suppresses residual echoes present in the recordings.
\item Denoise~\cite{zhao2022frcrn}: applies noise reduction techniques to further enhance the clarity and intelligibility of the speech.
\end{itemize}
This multi-stage cleaning pipeline ensures that background noises, reverberations, and other acoustic artifacts are minimized, significantly improving the quality of the training corpus.

\textbf{Data Formatting.}
After cleaning, we reprocess the audio to generate a high-quality parallel corpus. First, we segment the 1-minute chunks into individual sentences and discard segments shorter than 5 seconds. We use NVIDIA NeMo~\cite{nemo} for speaker diarization and retain only utterances containing a single speaker.
Each sentence is then scored using the MOS estimate provided by NISQA. We retain segments with MOS scores above 3.8, effectively filtering out approximately 30\% of the data. For transcription, we employ Whisper-large-v3 for English and FunASR~\cite{gao2023funasr} for Chinese.
The final dataset comprises over 100,000 hours of high-quality speech and corresponding transcriptions, forming a robust parallel corpus suitable for TTS training in long-form audio scenarios such as podcasts.

The full data processing pipeline is executed on a compute cluster equipped with NVIDIA A10 GPUs. The total processing required approximately 60,000 GPU hours, translating to a throughput of about 2.5 hours of raw audio per GPU hour.

\subsection{Pre-training of LLM}

As shown in Figure~\ref{fig:framework}, we continue pre-training Llama-3.2-3B on the collected parallel corpus, enabling it to model the relationship between text and discrete audio representations. The quantizer used in our framework generates 1024 unique audio tokens, necessitating the expansion of the LLM’s vocabulary with new tokens ranging from $\textless|audio\_token\_0|\textgreater$ to $\textless|audio\_token\_1023|\textgreater$, along with a special token $\textless|audio\_token\_end|\textgreater$ to signify the end of an audio sequence. The dataset is structured in an unsupervised training format, for example:
$$Hey,\ great\ to\ have\ you\ in\ Chatpods.\ \textless|audio\_token\_520|\textgreater...\textless|audio\_token\_end|\textgreater$$
Note that we do not introduce a specific “Turn of Speech” token between text and audio tokens, as the model naturally distinguishes between the modalities.

The pre-training process is conducted on a high-performance computing cluster comprising 80 NVIDIA A100 (80GB, NVLink) GPUs. Each compute node in the cluster contains 8 GPUs, interconnected via InfiniBand (IB) to facilitate efficient inter-node communication. The model is trained for 15 epochs with a learning rate of 1e-4, which takes approximately 10 days to complete.

Despite being trained in an unsupervised manner, the pre-trained LLM demonstrates strong capabilities for zero-shot voice cloning by leveraging its learned continuation abilities. During inference, we structure the input as a combination of reference\_text\_tokens + target\_text\_tokens + reference\_audio\_tokens and prompt the model to generate the corresponding target\_audio\_tokens. These generated tokens can then be decoded into speech that replicates the vocal characteristics of the reference audio while articulating the target text.

\subsection{Post-training of LLM}

While the pre-trained LLM exhibits zero-shot capabilities in speech synthesis, its performance can be significantly improved when adapting to the voice characteristics of a specific speaker. To this end, we perform post-training on individual speech data in a supervised manner.

We collect a dataset comprising several dozen minutes to a few hours of recordings from a single speaker. This data is then used to fine-tune the LLM, enabling it to generate more natural and personalized speech outputs that better reflect the target speaker's vocal identity.

The fine-tuning process follows a supervised training paradigm. Specifically, we employ an instruction-following format inspired by Alpaca~\cite{taori2023alpaca}. Each training sample consists of a textual prompt and its corresponding synthesized audio token sequence. An example of the dataset format is shown below:
\begin{lstlisting}
{
  "instruction": "Hey, great to have you in Chatpods.",
  "input": "",
  "output": "<|audio_token_520|>...<|audio_token_end|>"
}
\end{lstlisting}
The data samples are assembled using the \emph{llama3} template during post-training, and the same template is adopted at inference time with instruction-only input.

For each speaker's dataset, we post-train the LLM for 10 epochs using a learning rate of 1e-5. On a single node equipped with 8 NVIDIA A100 (40G, PCIe) GPUs, it takes approximately 15 minutes to train on one hour of speech data. This cost is negligible in comparison to the computational requirements of the pre-training stage.

\subsection{Training of Decoder}

The SoVITS model pre-trained in GPT-SoVITS can be directly used as the decoder to synthesize speech, as its quantizer is aligned with the training data of the language model. However, to optimize for the podcast scenario, we fine-tune the decoder on a curated subset of high-quality podcast audio.

To this end, we sample 10,000 hours of audio data from our dataset, selecting only clips with a MOS value greater than 4.5, as evaluated using NISQA. This ensures that the decoder is trained on speech data with high perceptual quality, improving synthesis fidelity and naturalness in the target domain.

The SoVITS model is trained (as referenced in~\cite{gptsovits}) for 8 epochs on the selected subset. The training process takes approximately one week using a single node equipped with 8 NVIDIA A100 GPUs (80GB, NVLink).

So far, we have presented the whole training process of Muyan-TTS, with a total budget of about \$50K. Table~\ref{tab:budget} presents an overview of the training costs associated with each major component.

\begin{table}[h]
    \footnotesize
    \centering
    \caption{Training costs of Muyan-TTS, assuming the rental price of A10 and A100 in GPU hour is \$0.5 and \$1, respectively.}
    \label{tab:budget}
    \begin{tabular}{c|c c c|c}
    \hline
        \textbf{Training Cost} & \textbf{Data Processing} & \textbf{Pre-training of LLM} & \textbf{Training of Decoder} & \textbf{Total} \\
    \hline
        in GPU Hours & 60K (A10) & 19.2K (A100) & 1.34K (A100) & - \\
    \hline
        in USD & \$30K & \$19.2K & \$1.34K & \$50.54K \\
    \hline
    \end{tabular}
\end{table}

\subsection{Inference Framework and Acceleration}

As discussed in the previous sections, the base model and the SFT model adopt different training objectives and input data formats, which lead to distinct inference modes. The base model requires both reference and target text tokens, along with reference audio tokens, as input and operates under an \emph{empty} template during inference. In contrast, the SFT model requires only the target text tokens as input, utilizing a structured \emph{llama3} template to guide generation.

To accelerate inference, we apply several optimization techniques, primarily targeting the LLM component. Efficient memory management strategies are employed, such as vLLM~\cite{kwon2023efficient}, which enables high-throughput decoding with minimal memory overhead. Additionally, we normalize the target text and perform sentence splitting to enable parallel inference, thereby improving latency and throughput across longer inputs.

\section{Experiments}

\subsection{Experimental Settings}

To comprehensively evaluate the performance of our proposed Muyan-TTS model, we conduct comparisons with several recent and competitive open-source TTS systems. These include CosyVoice2, Step-Audio, Spark-TTS~\cite{wang2025spark}, FireRedTTS~\cite{guo2024fireredtts}, and GPT-SoVITS v3. These baselines are selected for their state-of-the-art architectures and publicly available implementations, representing diverse design philosophies in modern TTS systems.

\textbf{Baseline Models}
\begin{itemize}[leftmargin=*]
    \item \textbf{CosyVoice2} integrates a pre-trained Qwen2.5-0.5B LLM as its backbone and adopts flow-matching for waveform synthesis. Its speech tokenizer is pre-trained on 200,000 hours of audio, while the full model is fine-tuned on an additional 160,000 hours of paired data.
    \item \textbf{Step-Audio} employs a more powerful 3B-parameter LLM decoder and also leverages flow-matching for waveform generation. The model is trained on an extensive dataset exceeding 1 million hours of speech data.
    \item \textbf{Spark-TTS} is built upon the same Qwen2.5-0.5B LLM backbone as CosyVoice2. However, instead of using flow-matching, it utilizes the BiCodec decoder for direct audio generation from the LLM output. Its training corpus includes over 100,000 hours of audio.
    \item \textbf{FireRedTTS} features a 400M-parameter LLM and employs flow-matching as its decoder mechanism. This model is trained on over 200,000 hours of speech data.
    \item \textbf{GPT-SoVITS v3} integrates a 330M-parameter LLM with a 77M-parameter SoVITS module for speech synthesis. Its training dataset consists of approximately 7,000 hours of speech recordings.
\end{itemize}
The key architectural details and training configurations of these models, along with those of our Muyan-TTS system, are summarized in Table~\ref{tab:tts_comparison}.

\begin{table}[h]
    \centering
    \caption{Comparison of Muyan-TTS with recent open-source TTS models. In CosyVoice2, ``(tok.)" and ``(mod.)" denote the speech tokenizer and the LLM model, respectively.}
    \label{tab:tts_comparison}
    \begin{tabular}{l c l c}
        \toprule
        \textbf{Model}         & \textbf{LLM Parameter}      & \textbf{Decoder} & \textbf{Training Data (Hours)} \\
        \midrule
        Muyan-TTS             & 3B                          & SoVITS-based             & 100,000+                        \\
        CosyVoice2             & 0.5B                        & Flow-matching            & 200,000 (tok.), 160,000 (mod.) \\
        Step-Audio             & 3B                          & Flow-matching            & 1,000,000+                     \\
        Spark-TTS              & 0.5B                        & BiCodec Decoder          & 100,000+                       \\
        FireRedTTS             & 400M                        & Flow-matching            & 200,000+                       \\
        GPT-SoVITS v3          & 330M                        & SoVITS-based             & 7,000+                          \\
        \bottomrule
    \end{tabular}
\end{table}

\textbf{Evaluation Datasets}

We evaluate all models on two widely-adopted public speech corpora:
\begin{itemize}[leftmargin=*]
    \item The \textbf{LibriSpeech} dataset~\cite{panayotov2015librispeech}, which contains clean and challenging test sets curated from public-domain audiobooks. It is widely used for both ASR and TTS evaluations.
    \item The \textbf{SEED} dataset~\cite{anastassiou2024seed}, a modern benchmark for personalized and expressive speech synthesis, offering high-quality multi-speaker prompts and transcripts.
\end{itemize}
These datasets allow for fair and consistent evaluation across different models and ensure that both transcription fidelity and speaker consistency are rigorously assessed.

\textbf{Evaluation Metrics}

We employ three commonly used and complementary metrics to assess the performance of TTS models: Word Error Rate (WER), Speaker Similarity (SIM), and Mean Opinion Score (MOS).
\begin{itemize}[leftmargin=*]
    \item \textbf{WER}: This metric measures the transcription accuracy of the synthesized speech using ASR. We adopt the Whisper-large v3 model as the ASR backend, which provides state-of-the-art transcription quality across diverse acoustic conditions.
    \item \textbf{SIM}: To assess the preservation of speaker identity, we compute cosine similarity between speaker embeddings extracted from the reference and synthesized speech. Embeddings are obtained using WavLM-large~\cite{chen2022large}, a model fine-tuned for speaker verification tasks. Higher SIM values indicate better alignment of speaker characteristics.
    \item \textbf{MOS}: The perceptual quality of the generated audio is estimated using the NISQA v2 model, a neural network-based framework for objective speech quality assessment. NISQA provides automated MOS predictions that closely correlate with human judgments, offering a scalable alternative to labor-intensive subjective evaluations.
\end{itemize}
All evaluations are conducted under consistent conditions and batch settings, ensuring reproducibility and fairness across model comparisons.

\subsection{Experimental Results}
\subsubsection{Zero-Shot TTS Synthesis}
We begin our evaluation by assessing the base model’s capability in zero-shot TTS synthesis. For the LibriSpeech dataset, we employ an out-of-domain speaker sample as the prompt speech and use its corresponding transcript as the prompt text to perform zero-shot TTS synthesis on the \textit{test-clean} subset. For the SEED dataset, which includes designated prompt speech and corresponding text, we utilize these provided pairs to perform TTS synthesis consistently across all compared models.
It is important to note that Step-Audio does not support explicit timbre control. Consequently, we use its default voice configuration, \textit{Tingting}, for synthesizing speech across both datasets.
The comparative performance results of all evaluated models are summarized in Table~\ref{tab:main_table}.

\begin{table}[h]
    \centering
    \caption{Performance comparison of different models on LibriSpeech and SEED datasets. }
    \label{tab:main_table}
    \small 
    \begin{tabular}{l *{3}{c} *{3}{c}}
        \toprule
        \textbf{Model}      & \multicolumn{3}{c}{\textbf{LibriSpeech}} & \multicolumn{3}{c}{\textbf{SEED}} \\
        \cmidrule(lr){2-4} \cmidrule(lr){5-7}
                            & \textbf{WER(\%)$\downarrow$} & \textbf{MOS$\uparrow$} & \textbf{SIM$\uparrow$} & \textbf{WER(\%)$\downarrow$} & \textbf{MOS$\uparrow$} & \textbf{SIM$\uparrow$} \\
        \midrule
        CosyVoice 2         & 2.91  & 4.81  & 0.70          & 2.98 & 4.22           & 0.66 \\
        Step-Audio         & 5.22           & 4.90             & --                      & 2.73             & 4.90             & --              \\
        Spark-TTS          & 27.36          & 3.66             & 0.45                   & 3.04          & 4.04           & 0.57 \\
        FireRedTTS          & 9.58          & 5.00           & 0.48         & 9.58             & 4.07             & 0.46              \\
        GPT-SoVITS v3       & 6.02           & 4.28           & 0.31                   & 4.74          & 3.86           & 0.51 \\
        
        Muyan-TTS          & 3.44           & 4.58           & 0.37                   & 4.09          & 4.32  & 0.41 \\
        \bottomrule
    \end{tabular}
\end{table}

On the LibriSpeech \textit{test-clean} set, our model demonstrates competitive performance with respect to WER, achieving the second-lowest WER among all models—only slightly behind CosyVoice2—and outperforming the remaining baselines. This outcome suggests that the VITS-based architecture employed by our model plays a key role in mitigating hallucination issues commonly observed when leveraging LLMs in TTS. In terms of MOS, Muyan-TTS surpasses both Spark-TTS and GPT-SoVITS v3, indicating that scaling up the training corpus and LLM parameters positively contributes to the perceptual quality of the synthesized speech.

On the SEED \textit{test-en} subset, Muyan-TTS achieves a lower WER than both FireRedTTS and GPT-SoVITS v3, while attaining the second-highest MOS, exceeded only by Step-Audio. However, it is critical to acknowledge that Step-Audio performs synthesis using only the text from the SEED test-en set and does not utilize the corresponding prompt speech. Taking this into account, among all models that use both text and speech prompts from the SEED dataset, Muyan-TTS achieves the highest MOS.

While Muyan-TTS exhibits strong performance in both intelligibility and perceptual quality, it does not obtain the best SIM scores across either dataset. This limitation stems from the fact that the base model is not explicitly optimized for the task of voice cloning during pre-training. We identify this as an area for future improvement and intend to explore targeted optimization strategies to enhance voice similarity in future iterations of the model.

\subsubsection{Comparison with Supervised Fine-Tuned Model}

Our base model demonstrates strong zero-shot TTS synthesis capabilities. To further adapt it to the voice characteristics of a specific target speaker, we sample several dozen minutes of speech from an out-of-domain speaker and apply SFT, resulting in Muyan-TTS-SFT. We then compare the performance of Muyan-TTS-SFT with the base model on the LibriSpeech \textit{test-clean} set. The results are summarized in Table~\ref{tab:base_vs_sft}.
To ensure alignment with the training data format used during SFT, we normalize the transcripts from LibriSpeech by capitalizing the first letter of each sentence and appending a full stop at the end of each utterance. This preprocessing step reflects the formatting used in fine-tuning and helps evaluate the model under realistic inference conditions.

\begin{table}[h]
    \centering
    \caption{Comparison of our base model and SFT model.} 
    \label{tab:base_vs_sft}
    \begin{tabular}{l|c c c } 
        \toprule
        \textbf{Model} & \textbf{WER(\%)$\downarrow$}   & \textbf{MOS$\uparrow$} & \textbf{SIM$\uparrow$}    \\ 
        \midrule
        Muyan-TTS & 3.44      & 4.58     &  0.37           \\ 
        Muyan-TTS-SFT & 4.48 & 4.97     &  0.46      \\  
        \bottomrule
    \end{tabular}
\end{table}

As shown in Table~\ref{tab:base_vs_sft}, Muyan-TTS-SFT outperforms the base model in both MOS and SIM, indicating enhanced speech quality and more accurate voice adaptation. These improvements validate the effectiveness of supervised fine-tuning for speaker adaptation in TTS. However, we observe a minor degradation in WER. This can be attributed to the increased sensitivity of the fine-tuned model to input formatting, particularly the expectation of a full stop at the end of the text. Such dependency reflects a trade-off introduced by the training strategy, where strict formatting improves synthesis quality but slightly limits generalization to text inputs outside the fine-tuning distribution.

\subsubsection{Synthesis Speed}

To evaluate the effectiveness of our proposed inference acceleration framework, we benchmark the synthesis speed of Muyan-TTS against several representative open-source TTS models.

We quantify synthesis speed using the ratio:
\begin{equation}
    r=\frac{T_{\text{inf}}}{T_{\text{syn}}}
\end{equation}
where $T_{\text{inf}}$ is the total inference time required to generate an utterance, and $T_{\text{syn}}$ is the duration (in seconds) of the resulting synthesized speech. A smaller $r$ value indicates higher synthesis efficiency, meaning less computation time is needed to generate each second of audio. This ratio is particularly informative for latency-sensitive use cases, as it directly reflects real-time synthesis capability.

All models are evaluated using their official inference implementations, ensuring that any optimizations or custom logic embedded in the original codebases are preserved. Furthermore, to ensure a fair and unbiased comparison, we adopt a consistent evaluation protocol across all systems:
\begin{itemize}[leftmargin=*]
    \item A shared dataset sampled from the LibriSpeech corpus is used as input.
    \item All models operate in non-streaming inference mode, avoiding any pipeline advantages associated with real-time chunked synthesis.
    \item Sentence splitting and parallelization mechanisms are explicitly disabled during synthesis, to focus solely on intrinsic model efficiency.
\end{itemize}

\begin{table}[h]
    \centering
    \caption{The inference speed of different open-source TTS models.} 
    \label{tab:speed}
    \small 
    \begin{tabular}{c|c c c c c c} 
        \toprule
        \textbf{Model} & CosyVoice2   & Step-Audio & Spark-TTS & FireRedTTS & GPT-SoVITS v3 & Muyan-TTS   \\ 
        \midrule
        $r\downarrow$ & 2.19      & 0.90    & 1.31 & 0.61 & 0.48 & 0.33            \\ 
        \bottomrule
    \end{tabular}
\end{table}

The results, presented in Table~\ref{tab:speed}, demonstrate that Muyan-TTS achieves the fastest synthesis speed among all tested models, requiring only 0.33 seconds to synthesize one second of speech. This performance highlights the model’s suitability for latency-sensitive applications and real-time speech generation scenarios.

\subsubsection{SoVITS Training}

While the performance of the LLM plays a crucial role in TTS synthesis, it does not solely determine the overall quality of the synthesized speech. The characteristics and training of the SoVITS component can also significantly impact final performance. As discussed in Section 3.5, we investigate how both the quality of training data and the number of training epochs affect the performance of SoVITS, independent of the LLM. To isolate these factors, we fix the LLM component to the best-performing Muyan-TTS-SFT model and train multiple SoVITS models under different conditions.
We construct two subsets of training data based on MOS thresholds: one containing utterances with MOS $>$ 3.8 and another with MOS $>$ 4.5. For each subset, we trained SoVITS models for 4 and 8 epochs. The results are summarized in Table~\ref{tab:sovits}.

\begin{table}[h]
    \centering
    \caption{Comparison of different SoVITS models.} 
    \label{tab:sovits}
    \begin{tabular}{l|c c c } 
        \toprule
        \textbf{Model} & \textbf{WER(\%)$\downarrow$}   & \textbf{MOS$\uparrow$} & \textbf{SIM$\uparrow$}    \\ 
        \midrule
         Muyan-TTS-SFT+SoVITS(MOS$>$3.8,epoch=4) & 6.91 & 4.92     & 0.40       \\  
         Muyan-TTS-SFT+SoVITS(MOS$>$3.8,epoch=8) & 6.11 & 4.93     & 0.35       \\  
         Muyan-TTS-SFT+SoVITS(MOS$>$4.5,epoch=4) & 5.83 & 4.92     & 0.43       \\  
         Muyan-TTS-SFT+SoVITS(MOS$>$4.5,epoch=8) & 4.48 & 4.97     & 0.46       \\ 
        \bottomrule
    \end{tabular}
\end{table}

The results indicate that both the quality of training data and the number of training epochs substantially affect model performance. SoVITS models trained on the higher-quality MOS$>$4.5 dataset consistently outperformed those trained on the MOS$>$3.8 dataset, particularly in WER and SIM. Notably, the best-performing configuration—Muyan-TTS-SFT+SoVITS(MOS$>$4.5,epoch=8)—achieved the lowest WER (4.48\%), highest MOS (4.97), and highest SIM (0.46), underscoring the benefits of high-quality training data.
Interestingly, increasing the number of epochs did not always lead to performance improvements. For the MOS$>$3.8 dataset, extending training to 8 epochs resulted in a decline in SIM from 0.40 to 0.35, suggesting that prolonged training on lower-quality data may hinder the model’s ability to capture speaker characteristics effectively. This contrast highlights that while training duration is important, the quality of data plays a more dominant role in SoVITS performance, particularly in producing natural, intelligible, and speaker-consistent speech.

\section{Conclusion}
In this work, we introduce Muyan-TTS, an open-source, trainable text-to-speech system optimized for podcast applications under a \$50,000 budget. By integrating an LLM with a VITS-based decoder, Muyan-TTS offers high-quality, zero-shot synthesis while maintaining robustness and inference efficiency. Our approach bridges the gap between end-to-end robustness and LLM-based expressiveness by aligning textual and acoustic modalities through quantized audio tokens.
We present a fully transparent training pipeline, including data collection from over 100,000 hours of podcast content, model pre-training and SFT, and decoder adaptation using high-MOS data. Experiments across standard benchmarks show that Muyan-TTS performs competitively with state-of-the-art open-source models in terms of intelligibility, naturalness, and speed, while maintaining flexibility for voice adaptation.
Our findings also demonstrate the importance of high-quality data and decoder refinement in achieving perceptually pleasing speech, and we show that speaker similarity can be improved through SFT. Furthermore, Muyan-TTS achieves the fastest inference speed among all compared models, making it a strong candidate for real-time applications.
By releasing our code, model checkpoints, and training procedures, we hope Muyan-TTS will serve as a practical foundation for future research and deployment in podcast and voice-interaction scenarios.

\section{Limitations}

Despite the promising performance of Muyan-TTS, several limitations remain that warrant future investigation. First, the model’s reliance on a G2P module within the decoder introduces a dependency on having access to the complete phoneme sequence of the input text prior to synthesis, which precludes the possibility of streaming inference. Second, Muyan-TTS demonstrates limited capabilities in multilingual speech synthesis, because the training data is heavily skewed toward English. Finally, Muyan-TTS does not currently support instruction-following TTS tasks due to the absence of instruction-level annotations in the training corpus.

\section*{Acknowledgment}

We would like to express our sincere gratitude to Shunshun Yin from Soul, Shengmin Jiang from Zhipu AI, and RVC-Boss, the author of GPT-SoVITS, for their invaluable support and insightful discussions from the beginning of the project.

\bibliographystyle{unsrt}
\bibliography{ref}

\begin{thebibliography}{10}

\bibitem{kim2021conditional}
Jaehyeon Kim, Jungil Kong, and Juhee Son.
\newblock Conditional variational autoencoder with adversarial learning for
  end-to-end text-to-speech.
\newblock In {\em International Conference on Machine Learning}, pages
  5530--5540. PMLR, 2021.

\bibitem{kong2023vits2}
Jungil Kong, Jihoon Park, Beomjeong Kim, Jeongmin Kim, Dohee Kong, and Sangjin
  Kim.
\newblock Vits2: Improving quality and efficiency of single-stage
  text-to-speech with adversarial learning and architecture design.
\newblock {\em arXiv preprint arXiv:2307.16430}, 2023.

\bibitem{weiss2021wave}
Ron~J Weiss, RJ~Skerry-Ryan, Eric Battenberg, Soroosh Mariooryad, and
  Diederik~P Kingma.
\newblock Wave-tacotron: Spectrogram-free end-to-end text-to-speech synthesis.
\newblock In {\em ICASSP 2021-2021 IEEE International Conference on Acoustics,
  Speech and Signal Processing (ICASSP)}, pages 5679--5683. IEEE, 2021.

\bibitem{donahue2020end}
Jeff Donahue, Sander Dieleman, Miko{\l}aj Bi{\'n}kowski, Erich Elsen, and Karen
  Simonyan.
\newblock End-to-end adversarial text-to-speech.
\newblock {\em arXiv preprint arXiv:2006.03575}, 2020.

\bibitem{ping2018clarinet}
Wei Ping, Kainan Peng, and Jitong Chen.
\newblock Clarinet: Parallel wave generation in end-to-end text-to-speech.
\newblock {\em arXiv preprint arXiv:1807.07281}, 2018.

\bibitem{gptsovits}
RVC-Boss.
\newblock Gpt-sovits.
\newblock \url{https://github.com/RVC-Boss/GPT-SoVITS}, 2024.

\bibitem{chattts}
2Noise.
\newblock Chattts.
\newblock \url{https://github.com/2noise/ChatTTS}, 2024.

\bibitem{li2019neural}
Naihan Li, Shujie Liu, Yanqing Liu, Sheng Zhao, and Ming Liu.
\newblock Neural speech synthesis with transformer network.
\newblock In {\em Proceedings of the AAAI conference on artificial
  intelligence}, volume~33, pages 6706--6713, 2019.

\bibitem{shen2018natural}
Jonathan Shen, Ruoming Pang, Ron~J Weiss, Mike Schuster, Navdeep Jaitly,
  Zongheng Yang, Zhifeng Chen, Yu~Zhang, Yuxuan Wang, Rj~Skerrv-Ryan, et~al.
\newblock Natural tts synthesis by conditioning wavenet on mel spectrogram
  predictions.
\newblock In {\em 2018 IEEE international conference on acoustics, speech and
  signal processing (ICASSP)}, pages 4779--4783. IEEE, 2018.

\bibitem{valle2020flowtron}
Rafael Valle, Kevin Shih, Ryan Prenger, and Bryan Catanzaro.
\newblock Flowtron: an autoregressive flow-based generative network for
  text-to-speech synthesis.
\newblock {\em arXiv preprint arXiv:2005.05957}, 2020.

\bibitem{wang2023neural}
Chengyi Wang, Sanyuan Chen, Yu~Wu, Ziqiang Zhang, Long Zhou, Shujie Liu, Zhuo
  Chen, Yanqing Liu, Huaming Wang, Jinyu Li, et~al.
\newblock Neural codec language models are zero-shot text to speech
  synthesizers.
\newblock {\em arXiv preprint arXiv:2301.02111}, 2023.

\bibitem{chen2024vall}
Sanyuan Chen, Shujie Liu, Long Zhou, Yanqing Liu, Xu~Tan, Jinyu Li, Sheng Zhao,
  Yao Qian, and Furu Wei.
\newblock Vall-e 2: Neural codec language models are human parity zero-shot
  text to speech synthesizers.
\newblock {\em arXiv preprint arXiv:2406.05370}, 2024.

\bibitem{du2023lauragpt}
Zhihao Du, Jiaming Wang, Qian Chen, Yunfei Chu, Zhifu Gao, Zerui Li, Kai Hu,
  Xiaohuan Zhou, Jin Xu, Ziyang Ma, et~al.
\newblock Lauragpt: Listen, attend, understand, and regenerate audio with gpt.
\newblock {\em arXiv preprint arXiv:2310.04673}, 2023.

\bibitem{liao2024fish}
Shijia Liao, Yuxuan Wang, Tianyu Li, Yifan Cheng, Ruoyi Zhang, Rongzhi Zhou,
  and Yijin Xing.
\newblock Fish-speech: Leveraging large language models for advanced
  multilingual text-to-speech synthesis.
\newblock {\em arXiv preprint arXiv:2411.01156}, 2024.

\bibitem{anastassiou2024seed}
Philip Anastassiou, Jiawei Chen, Jitong Chen, Yuanzhe Chen, Zhuo Chen, Ziyi
  Chen, Jian Cong, Lelai Deng, Chuang Ding, Lu~Gao, et~al.
\newblock Seed-tts: A family of high-quality versatile speech generation
  models.
\newblock {\em arXiv preprint arXiv:2406.02430}, 2024.

\bibitem{du2024cosyvoice1}
Zhihao Du, Qian Chen, Shiliang Zhang, Kai Hu, Heng Lu, Yexin Yang, Hangrui Hu,
  Siqi Zheng, Yue Gu, Ziyang Ma, et~al.
\newblock Cosyvoice: A scalable multilingual zero-shot text-to-speech
  synthesizer based on supervised semantic tokens.
\newblock {\em arXiv preprint arXiv:2407.05407}, 2024.

\bibitem{neekhara2024improving}
Paarth Neekhara, Shehzeen Hussain, Subhankar Ghosh, Jason Li, Rafael Valle,
  Rohan Badlani, and Boris Ginsburg.
\newblock Improving robustness of llm-based speech synthesis by learning
  monotonic alignment.
\newblock {\em arXiv preprint arXiv:2406.17957}, 2024.

\bibitem{xue2024improving}
Jinlong Xue, Yayue Deng, Yicheng Han, Yingming Gao, and Ya~Li.
\newblock Improving audio codec-based zero-shot text-to-speech synthesis with
  multi-modal context and large language model.
\newblock {\em arXiv preprint arXiv:2406.03706}, 2024.

\bibitem{du2024cosyvoice}
Zhihao Du, Yuxuan Wang, Qian Chen, Xian Shi, Xiang Lv, Tianyu Zhao, Zhifu Gao,
  Yexin Yang, Changfeng Gao, Hui Wang, et~al.
\newblock Cosyvoice 2: Scalable streaming speech synthesis with large language
  models.
\newblock {\em arXiv preprint arXiv:2412.10117}, 2024.

\bibitem{huang2025step}
Ailin Huang, Boyong Wu, Bruce Wang, Chao Yan, Chen Hu, Chengli Feng, Fei Tian,
  Feiyu Shen, Jingbei Li, Mingrui Chen, et~al.
\newblock Step-audio: Unified understanding and generation in intelligent
  speech interaction.
\newblock {\em arXiv preprint arXiv:2502.11946}, 2025.

\bibitem{grattafiori2024llama}
Aaron Grattafiori, Abhimanyu Dubey, Abhinav Jauhri, Abhinav Pandey, Abhishek
  Kadian, Ahmad Al-Dahle, Aiesha Letman, Akhil Mathur, Alan Schelten, Alex
  Vaughan, et~al.
\newblock The llama 3 herd of models.
\newblock {\em arXiv preprint arXiv:2407.21783}, 2024.

\bibitem{yang2024qwen2}
An~Yang, Baosong Yang, Beichen Zhang, Binyuan Hui, Bo~Zheng, Bowen Yu,
  Chengyuan Li, Dayiheng Liu, Fei Huang, Haoran Wei, et~al.
\newblock Qwen2. 5 technical report.
\newblock {\em arXiv preprint arXiv:2412.15115}, 2024.

\bibitem{radford2023robust}
Alec Radford, Jong~Wook Kim, Tao Xu, Greg Brockman, Christine McLeavey, and
  Ilya Sutskever.
\newblock Robust speech recognition via large-scale weak supervision.
\newblock In {\em International conference on machine learning}, pages
  28492--28518. PMLR, 2023.

\bibitem{hsu2021hubert}
Wei-Ning Hsu, Benjamin Bolte, Yao-Hung~Hubert Tsai, Kushal Lakhotia, Ruslan
  Salakhutdinov, and Abdelrahman Mohamed.
\newblock Hubert: Self-supervised speech representation learning by masked
  prediction of hidden units.
\newblock {\em IEEE/ACM transactions on audio, speech, and language
  processing}, 29:3451--3460, 2021.

\bibitem{mittag2021nisqa}
Gabriel Mittag, Babak Naderi, Assmaa Chehadi, and Sebastian M{\"o}ller.
\newblock Nisqa: A deep cnn-self-attention model for multidimensional speech
  quality prediction with crowdsourced datasets.
\newblock {\em arXiv preprint arXiv:2104.09494}, 2021.

\bibitem{lu2024music}
Wei-Tsung Lu, Ju-Chiang Wang, Qiuqiang Kong, and Yun-Ning Hung.
\newblock Music source separation with band-split rope transformer.
\newblock In {\em ICASSP 2024-2024 IEEE International Conference on Acoustics,
  Speech and Signal Processing (ICASSP)}, pages 481--485. IEEE, 2024.

\bibitem{liu2021voicefixer}
Haohe Liu, Qiuqiang Kong, Qiao Tian, Yan Zhao, DeLiang Wang, Chuanzeng Huang,
  and Yuxuan Wang.
\newblock Voicefixer: Toward general speech restoration with neural vocoder.
\newblock {\em arXiv preprint arXiv:2109.13731}, 2021.

\bibitem{yang2023low}
Dong Yang, Fei Jiang, Wei Wu, Xuefei Fang, and Muyong Cao.
\newblock Low-complexity acoustic echo cancellation with neural kalman
  filtering.
\newblock In {\em ICASSP 2023-2023 IEEE International Conference on Acoustics,
  Speech and Signal Processing (ICASSP)}, pages 1--5. IEEE, 2023.

\bibitem{zhao2022frcrn}
Shengkui Zhao, Bin Ma, Karn~N Watcharasupat, and Woon-Seng Gan.
\newblock Frcrn: Boosting feature representation using frequency recurrence for
  monaural speech enhancement.
\newblock In {\em ICASSP 2022-2022 IEEE international conference on acoustics,
  speech and signal processing (ICASSP)}, pages 9281--9285. IEEE, 2022.

\bibitem{nemo}
NVIDIA.
\newblock Nemo.
\newblock \url{https://github.com/NVIDIA/NeMo}, 2024.

\bibitem{gao2023funasr}
Zhifu Gao, Zerui Li, Jiaming Wang, Haoneng Luo, Xian Shi, Mengzhe Chen, Yabin
  Li, Lingyun Zuo, Zhihao Du, Zhangyu Xiao, et~al.
\newblock Funasr: A fundamental end-to-end speech recognition toolkit.
\newblock {\em arXiv preprint arXiv:2305.11013}, 2023.

\bibitem{taori2023alpaca}
Rohan Taori, Ishaan Gulrajani, Tianyi Zhang, Yann Dubois, Xuechen Li, Carlos
  Guestrin, Percy Liang, and Tatsunori~B Hashimoto.
\newblock Alpaca: A strong, replicable instruction-following model.
\newblock {\em Stanford Center for Research on Foundation Models.
  https://crfm.stanford.edu/2023/03/13/alpaca.html}, 3(6):7, 2023.

\bibitem{kwon2023efficient}
Woosuk Kwon, Zhuohan Li, Siyuan Zhuang, Ying Sheng, Lianmin Zheng, Cody~Hao Yu,
  Joseph Gonzalez, Hao Zhang, and Ion Stoica.
\newblock Efficient memory management for large language model serving with
  pagedattention.
\newblock In {\em Proceedings of the 29th Symposium on Operating Systems
  Principles}, pages 611--626, 2023.

\bibitem{wang2025spark}
Xinsheng Wang, Mingqi Jiang, Ziyang Ma, Ziyu Zhang, Songxiang Liu, Linqin Li,
  Zheng Liang, Qixi Zheng, Rui Wang, Xiaoqin Feng, et~al.
\newblock Spark-tts: An efficient llm-based text-to-speech model with
  single-stream decoupled speech tokens.
\newblock {\em arXiv preprint arXiv:2503.01710}, 2025.

\bibitem{guo2024fireredtts}
Hao-Han Guo, Kun Liu, Fei-Yu Shen, Yi-Chen Wu, Feng-Long Xie, Kun Xie, and
  Kai-Tuo Xu.
\newblock Fireredtts: A foundation text-to-speech framework for industry-level
  generative speech applications.
\newblock {\em arXiv preprint arXiv:2409.03283}, 2024.

\bibitem{panayotov2015librispeech}
Vassil Panayotov, Guoguo Chen, Daniel Povey, and Sanjeev Khudanpur.
\newblock Librispeech: an asr corpus based on public domain audio books.
\newblock In {\em 2015 IEEE international conference on acoustics, speech and
  signal processing (ICASSP)}, pages 5206--5210. IEEE, 2015.

\bibitem{chen2022large}
Zhengyang Chen, Sanyuan Chen, Yu~Wu, Yao Qian, Chengyi Wang, Shujie Liu, Yanmin
  Qian, and Michael Zeng.
\newblock Large-scale self-supervised speech representation learning for
  automatic speaker verification.
\newblock In {\em ICASSP 2022-2022 IEEE International Conference on Acoustics,
  Speech and Signal Processing (ICASSP)}, pages 6147--6151. IEEE, 2022.

\end{thebibliography}

\appendix

\end{document}